\documentclass[12pt,letter]{article}

\usepackage{subfigure}
\usepackage{graphicx, epsfig, color}
\usepackage{amsmath,amssymb}
\def\dsl{\partial \llap/}

\def\to{\rightarrow}

\def\bi{\begin{itemize}}
\def\ei{\end{itemize}}

\def\ta{\tilde a}
\def\tG{\tilde G}

\def\sps1ap{SPS1a$^\prime$}
\def\c1p{C1$^\prime$}

\def\tg{\tilde g}

\def\tw{\widetilde W}
\def\tz{\widetilde Z}
\def\alt{\lesssim}
\def\agt{\gtrsim}
\def\be{\begin{equation}}  
\def\ee{\end{equation}}  
\def\bea{\begin{eqnarray}}  
\def\eea{\end{eqnarray}}  
\def\beas{\begin{eqnarray*}}  
\def\eeas{\end{eqnarray*}}  
\newcommand\prd[3]{{\it Phys.\ Rev.\ }{\bf D #1} (#2) #3}

\newcommand\prl[3]{{\it Phys.\ Rev.\ Lett.\ }{\bf #1} (#2) #3}
\newcommand\plb[3]{{\it Phys.\ Lett.\ }{\bf B #1} (#2) #3}
\newcommand\jhep[3]{{\it J. High Energy Phys.\ }{\bf #1} (#2) #3}

\newcommand\npb[3]{{\it Nucl.\ Phys.\ }{\bf B #1} (#2) #3}

\begin{document}
\begin{titlepage}
\begin{flushright}
CTPU-17-16\\
OU-HEP-170421
\end{flushright}

\vspace{0.5cm}
\begin{center}
{\Large \bf Prospects for axion detection in natural SUSY
with mixed axion-higgsino dark matter:
back to invisible?
}\\ 
\vspace{1.2cm} \renewcommand{\thefootnote}{\fnsymbol{footnote}}
{\large Kyu Jung Bae$^{1}$\footnote[1]{Email: kyujungbae@ibs.re.kr},
Howard Baer$^{2}$\footnote[2]{Email: baer@nhn.ou.edu },
and Hasan Serce$^{2}$\footnote[3]{Email: serce@ou.edu } 
}\\ 
\vspace{1.2cm} \renewcommand{\thefootnote}{\arabic{footnote}}
{\it 
$^1$Center for Theoretical Physics of the Universe,\\
Institute for Basic Science (IBS), 
Daejeon 34051, Korea\\
}
{\it 
$^2$Dept. of Physics and Astronomy,\\
University of Oklahoma, Norman, OK 73019, USA \\
}

\end{center}

\vspace{0.5cm}
\begin{abstract}
\noindent 
Under the expectation that nature is natural, 
we extend the Standard Model to include SUSY to stabilize the 
electroweak sector and PQ symmetry to stabilize the QCD sector. 
Then natural SUSY arises from a Kim-Nilles solution to the 
SUSY $\mu$ problem which allows for a little hierarchy where 
$\mu\sim f_a^2/M_P\sim 100-300$ GeV while the SUSY particle 
mass scale $m_{\rm SUSY}\sim 1-10$ TeV $\gg \mu$. 
Dark matter then consists of two particles: 
a higgsino-like WIMP and a SUSY DFSZ axion. 
The range of allowed axion mass values $m_a$ depends on the mixed
axion-higgsino relic density. The range of $m_a$ is actually 
restricted in this case by limits on WIMPs from direct and indirect detection
experiments.
We plot the expected axion detection rate at microwave cavity experiments. 
The axion-photon-photon coupling is severely diminished by charged higgsino contributions
to the anomalous coupling. In this case, the axion may retreat, 
at least temporarily, back into the regime of near invisibility.
From our results, we urge new ideas for techniques which probe 
both deeper and more broadly into axion coupling versus axion 
mass parameter space.
\vspace*{0.8cm}

%\noindent PACS numbers: 12.60.Jv,14.80.Va,14.80.Ly

\end{abstract}

\end{titlepage}

\section{Introduction}
\label{sec:intro}

It seems to be a tautology that nature is natural~\cite{nat}. 
In particle physics, 
a working definition of naturalness is that each independent contribution 
to an observable ought to be comparable to or less than its measured value. 
For if one
contribution were far greater, then some other supposedly unrelated 
contribution would need to be fine-tuned to exactly the right 
opposite-sign value such as to maintain the measured value. 
Such a situation is considered highly implausible or unnatural. 
In many circumstances, naturalness
has turned out to be a reliable guide towards the correct laws of 
nature~\cite{review} while the presence of fine-tuning acts as a sieve to
filter out faulty theories.

An example from the QCD sector of the Standard Model (SM) 
arises from 't Hooft's~\cite{tHooft} solution to the 
old $U(1)_A$ problem~\cite{u1a} via the discovery of the $\theta$ vacuum.
A consequence of the instanton-induced $\theta$ vacuum is that the QCD
Lagrangian should contain an additional term
\begin{equation}
{\cal L}_{\rm QCD}\ni\frac{\alpha_s\bar{\theta}}{8\pi}G_{\mu\nu A}\tilde{G}^{\mu\nu}_A
\end{equation}
(where $G_{\mu\nu A}$ is the gluon field strength tensor)
which gives rise to strong $CP$ violating interactions. 
The term $\bar{\theta}$ contains two separate contributions 
$\bar{\theta}=\theta_{\rm QCD}+{\rm Arg}[{\rm det}(m_q)]$. 
Measurements of the neutron EDM require $\bar{\theta}\alt 10^{-10}$. 
The strong CP problem of QCD-- why $\bar{\theta}$ is so small-- 
is thus a problem of naturalness. So far, the most compelling solution 
to the strong CP problem is to impose an additional 
(spontaneously broken) global Peccei-Quinn (PQ)
symmetry which requires the existence of an (invisible) 
axion~\cite{pq,ww,ksvz,dfsz,axreview}.

In the electroweak (EW) sector of the SM, it is well known that
the Higgs boson mass is quadratically unstable under quantum corrections.
Increasingly implausible fine-tunings are necessary to maintain
the Higgs mass at its measured value $m_h\simeq 125$ GeV~\cite{atlas_h,cms_h} 
depending on the cut-off scale $\Lambda$ which demarcates the upper 
energy range of validity of the theory. 
A simple and elegant solution to the Higgs naturalness problem is
to extend the underlying Poincar\'e spacetime symmetries to include their 
maximal structure: supersymmetry (SUSY). 
The minimal supersymmetrized Standard Model,
or MSSM~\cite{wss}, 
is free of quadratic divergences but phenomenologically requires 
inclusion of soft SUSY breaking terms not too far from the weak scale.
The MSSM predicts the existence of a panoply 
of new superpartner matter states: squarks, sleptons, gluinos, 
charginos and neutralinos.
So far, superpartners have yet to be found at LHC~\cite{atlas_s,cms_s}; 
this has led to concern that the fine-tuning may creep back into the MSSM 
via log instead of quadratic divergences~\cite{nat}. 
A Little Hierarchy problem (LHP) has emerged~\cite{barbstrum}: 
how can the weak scale as characterized by $m_{W,Z,h}\sim 100$ GeV 
be stable while superpartners apparently lie in the multi-TeV range?

Some perspective on the LHP can be gained by examining the 
scalar potential minimization conditions which relate the measured value of
$m_Z$ to MSSM Lagrangian parameters:
\bea
\frac{m_Z^2}{2}&=&\frac{m_{H_d}^2+\Sigma_d^d-(m_{H_u}^2+\Sigma_u^u)\tan^2\beta}{\tan^2\beta -1}-\mu^2 \label{eq:mzs1} \\
&\simeq& -m_{H_u}^2-\Sigma_u^u-\mu^2,
\label{eq:mzs2}
\eea
where the latter approximate equality arises for $\tan\beta \agt 3$. 
The $m_{H_{u,d}}^2$ are weak scale soft SUSY breaking Higgs mass terms,
$\mu$ is the superpotential Higgs/higgsino mass parameter and 
$\tan\beta\equiv v_u/v_d$ is the ratio of Higgs field vevs. The terms 
$\Sigma_u^u$ and $\Sigma_d^d$ contain an assortment of 1-loop corrections;
expressions can be found in the Appendix of Ref.~\cite{rns}.
For weak scale naturalness with less than 3\% fine-tuning, from Eq. \ref{eq:mzs1} we see that
\begin{enumerate}
\item $\mu\sim 100-350$ GeV~\cite{mu} (the lower bound arises due to LEP2 searches for charginos),
\item $m_{H_u}^2\sim -(100-350)^2\ {\rm GeV}^2$ and
\item $\Sigma_u^u(i)\sim (100-350)^2$ GeV$^2$ 
\end{enumerate} 
(where $i$ labels the various loop contributions to $\Sigma$).
In the case of $m_{H_u}^2$: even if soft terms are at multi-TeV level 
at energy scale $Q\sim m_{\rm GUT}$, $m_{H_u}^2$ can be driven to natural 
values~\cite{ltr} at the weak scale thanks to the large top Yukawa coupling. 
This situation is known as radiatively driven naturalness.
The $\Sigma_u^u$ term is typically dominated by the top squark contributions. 
It is minimized for TeV-scale highly mixed top-squarks-- which are just the 
right conditions to lift $m_h$ up to $\sim 125$ GeV~\cite{h125}. 
Mass limits on sparticles from requiring 
\be
\Delta_{\rm EW}\equiv |\text{max term on RHS of Eq.}~\ref{eq:mzs1}|/(m_Z^2/2)<30
\ee
allow for $m_{\tg}\alt 5-6$ TeV so SUSY maintains naturalness even in the face
of LHC measurements of $m_h$ and limits on sparticle masses~\cite{rns,upper}.

An important consequence of $\mu\sim 100-350$ GeV is that 
light higgsinos $\tw_1^\pm$ and $\tz_{1,2}$ should exist. 
The lightest SUSY particle $\tz_1$ is a higgsino-like neutralino
which is thermally underproduced as dark matter~\cite{bbm}. 
In our setup, where we also require an axion to solve the strong CP problem, then axions also make up a
(potentially dominant) portion of the dark matter. 
The axion field is now just one element of an axion supermultiplet
\be
A=(s+ia)/\sqrt{2}+\sqrt{2}\theta\tilde{a}+\theta^2{\cal F}_a
\ee
where $\theta$ now are superspace co-ordinates, $s$ is the 
$R$-parity-even spin-0 
{\it saxion} field and $\tilde{a}$ is the $R$-parity-odd spin-$1/2$ {\it axino} field
and ${\cal F}_a$ is the auxiliary field of the axion supermultiplet.
In gravity-mediated SUSY breaking, one expects $m_s$ of order the 
gravitino mass $m_{3/2}\sim$ TeV (similarly to other scalar masses).
The axino mass $m_{\ta}$ is also typically of the same order ($\sim m_{3/2}$) 
although it can be different in some models~\cite{axnmass1,axnmass2,cl,bbcs15}. 
In this case, the higgsino-like WIMPs can be non-thermally produced~\cite{kimrev} 
due to production and late decays of axinos and saxions in the 
early universe~\cite{az1,chun}.

An important element of the LHP is to try to understand why 
$\mu\sim m_{W,Z,h}\ll m_{\rm SUSY}\sim 1-10$ TeV. 
The SUSY version of the DFSZ~\cite{dfsz}
axion model offers an elegant solution known as the Kim-Nilles (KN) 
mechanism~\cite{kn}. 
In KN, the Higgs and matter superfields actually carry PQ charge assignments
with $Q_{\rm PQ}(H_{u})+Q_{\rm PQ}(H_{d})=-(n+1)$ (and various possibilities occur for PQ charge assignments for matter superfields).
%\footnote{Once PQ charge assignments of Higgs superfields are set, physical 1PI intereactions are independent of PQ charge assigments for matter superfields~\cite{Bae:2011jb}.}
In this case, the $\mu$ term in the superpotential is forbidden which explains
why $\mu$ is not of order the reduced Planck mass 
$M_P\sim 2.4\times 10^{18}$ GeV. 
However, now one may introduce a SM singlet but PQ charged chiral superfield
$S$ with $Q_{\rm PQ}(S)=+1$ with non-renormalizable superpotential
\be
W\ni \lambda_\mu S^{n+1}H_u H_d/M_P^n .
\ee
Once the PQ symmetry is spontaneously broken, then the PQ fields 
develop a vev $\langle S\rangle\sim f_a$ where $f_a$ is the axion decay
constant and sets the PQ scale. A $\mu$ term develops with 
\be 
\mu\sim\lambda_\mu f_a^{n+1}/M_P^n .
\ee
For $\mu\sim 100$ GeV,  $n=1$ and $\lambda_\mu\sim 1$, 
then one expects $f_a\sim 10^{10}$ GeV. For higher $n$ values, then
much larger values of $f_a$ are expected. The value of $\mu$ here may be 
compared with the expected value of gravitino mass from simple 
supergravity models: $m_{3/2}\sim m_{\rm hidden}^2/M_P$ 
(where $m_{\rm hidden}$ denotes the mass scale associated with
the hidden sector). 
Then the Little Hierarchy $\mu\ll m_{3/2}$ develops as a consequence of 
$f_a\ll m_{\rm hidden}$.
In fact, there exists a class of models where PQ symmetry is broken radiatively
as a consequence of SUSY breaking. In these radiatively-broken  PQ models, it
is typical to develop $\mu\sim 100$ GeV as a consequence of 
$m_{\rm SUSY}\sim 1-10$ TeV~\cite{radpq}.

In previous papers~\cite{Bae:2015jea,wimp}, 
we have examined prospects for direct and indirect 
detection of the higgsino-like WIMPs from mixed axion-higgsino dark matter
as expected in natural SUSY models. In the present paper, we examine
prospects for axion detection at microwave cavity detectors such as 
ADMX~\cite{admx,rybka}. As with WIMP detection, our answer depends now not only on the axion mass and coupling, but also on the relative portion of dark matter
made up of axions vis-a-vis WIMPs. As such, in Sec. \ref{sec:rd}, 
we examine the WIMP vs. axion relic density in an updated natural SUSY 
benchmark model. An issue which arises anew here is that, for a particular set
of PQ parameters, if the WIMP abundance is enhanced compared to axions, 
then the particular parameter set may become ruled out by WIMP
direct or indirect detection search limits. 
In Sec. \ref{sec:axion}, we examine prospects for axion detection in the 
SUSY DFSZ model at microwave cavity experiments. We re-evaluate the
axion coupling $g_{a\gamma\gamma}$ and find that it is severely diminished
from the non-SUSY DFSZ model or the KSVZ model due to the circulation 
of PQ-charged higgsinos in the loop. Even so, a rather large range
of $m_a$ values emerge as possible search targets. 
The range is somewhat disjoint due to the interplay between non-thermal
WIMP production via axino or saxion production and decay.
In Sec. \ref{sec:conclude}, we summarize and present conclusions. 
In the SUSY DFSZ model, which we feel is most compelling due to 
simultaneously solving 1. the gauge hierarchy problem, 2. the strong 
CP problem, 3. the Little Hierarchy problem and 4. the SUSY $\mu$ problem, 
the discovery of the axion may require much deeper probes of 
axion coupling and much broader scans over axion mass. If not, the
axion may retreat back into invisibility, at least in the near term future.

\section{Relic density of axions and higgsino-like WIMPs}
\label{sec:rd}

In this Section, we compute the relic abundance of higgsino-like WIMPs 
and DFSZ axions from an updated natural SUSY benchmark model. 
An older SUSY benchmark model dubbed SUA in Ref.~\cite{dfsz2} 
with $m_{\tg}=1.8$ TeV has apparently been excluded by recent 
LHC searches which now require $m_{\tg}\agt 2$ TeV~\cite{atlas_s,cms_s}. 
The new natural SUSY benchmark point from the two-extra-parameter
non-universal Higgs model~\cite{nuhm2} (NUHM2) has parameters 
$m_0=5.3$ TeV, $m_{1/2}=2.03$ TeV, $A_0=-9.85$ TeV, $\tan\beta =9$ with
$(\mu ,m_A)=(150,3000)$ GeV. This gives rise to a gluino mass
$m_{\tg}=4.48$ TeV, well above the reach of HL-LHC but within reach of a
33 TeV LHC energy upgrade~\cite{lhc33}. The value of $\Delta_{\rm EW}=29$ for
$3.3\%$ EW fine-tuning. The thermally produced higgsino relic abundance is
$\Omega_{\tz_1}^{\rm TP}h^2=0.006$ or about 5\% of the measured dark matter abundance.
The higgsino-like WIMP has mass $m_{\tz_1}=150.4$ GeV.

To evaluate the mixed neutralino-axion relic density, we apply the
eight-coupled-Boltzmann equation computer code developed in 
Ref's~\cite{dfsz2,andre}. 
Starting from the time of re-heat with temperature $T_R$
at the end of the inflationary epoch, 
the computer code tracks the coupled abundances
of radiation ({\it i.e.} SM particles), neutralinos, axinos, gravitinos, 
saxions and axions (the latter two consist of thermal/decay-produced components
and coherent oscillations (CO)).
%are produced both thermally and 
%via coherent oscillations (CO)). 

The CO abundance of axions is determined by its initial misalignment angle $\theta_i$
~\cite{axdm,vg1}.
For numerical analyses, we adopt a simple formula
%and is given by
%
\be
\Omega_a^{\rm CO}h^2\simeq 0.23 f(\theta_i )\theta_i^2
\left(\frac{f_a/N_{\rm DW}}{10^{12}\ {\rm GeV}}\right)^{7/6}
\ee 
where $f(\theta_i )=\left[\log \left((e/(1-\theta_i^2/\pi^2)\right)\right]^{7/6}$
is the anharmonicity factor~\cite{vg1}. 
Provided the neutralino and thermal/decay-produced axion abundance is below 
the total measured DM abundance, the value of $\theta_i$ can always be 
adjusted so that CO-produced axions make up the remainder.
\begin{figure}[tbp]
\begin{center}
\includegraphics[height=0.35\textheight]{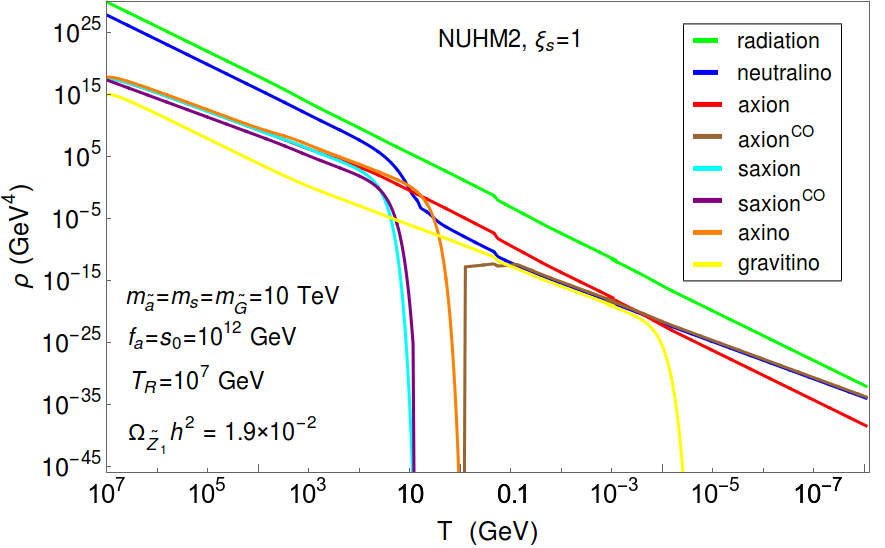}\\
\caption{A plot of various energy densities $\rho$ vs. 
temperature $T$ starting from $T_R=10^7$ GeV until the era of entropy conservation from our eight-coupled Boltzmann equation solution to the mixed
axion-neutralino relic density in the SUSY DFSZ model for 
a natural SUSY benchmark point. We take $\xi_s=1$.
\label{fig:rhoT}}
\end{center}
\end{figure}

A plot of the energy densities vs. $T$ is shown for the SUSY DFSZ 
axion model for our natural SUSY NUHM2 benchmark model is shown in Fig \ref{fig:rhoT}.
We take $T_R=10^7$ GeV \footnote{This value of $T_R$
is in accord with baryogenesis mechanisms such as non-thermal or Affleck-Dine
leptogenesis~\cite{lepto}.} and $f_a=s_0=10^{12}$ GeV. 
We also take $m_{\ta}=m_s=m_{3/2}=10$ TeV.
The blue curve denotes the neutralino abundance which freezes out at
$T\sim 10$ GeV. Saxions decay around $T\sim 10$ GeV whilst axinos decay around $T\sim 1$ GeV, the temperature also where axions start to oscillate. 
Due to late decay of axinos which occurs after the freeze-out, the neutralino abundance increases to $\Omega_{\tz_1}h^2 \simeq 0.019.$
Next, we scan over the following PQMSSM parameter ranges:
\bi
\item $f_a:\ 10^9-10^{16}$ GeV,
\item $m_s:\ 1-40$ TeV,
\item $m_{\ta}:\ 1-40$ TeV,
\ei
while keeping the gravitino mass fixed at $m_{\tG}=10$ TeV and the reheat temperature fixed at $T_R=10^7$ GeV.
\begin{figure}
\begin{center}
\subfigure[]
{\includegraphics[width=8.4cm]{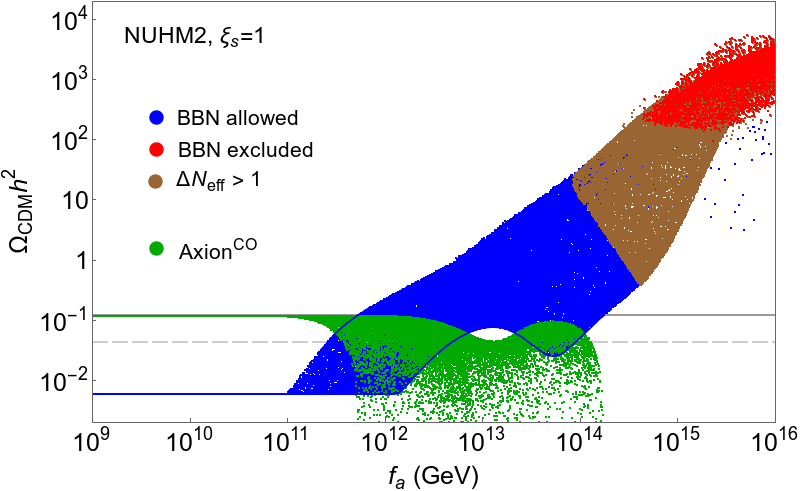}}
%{\includegraphics[width=14cm]{nuhm2_DM.png}}
\subfigure[]
{%\hspace*{.11in}
\includegraphics[width=8.4cm]{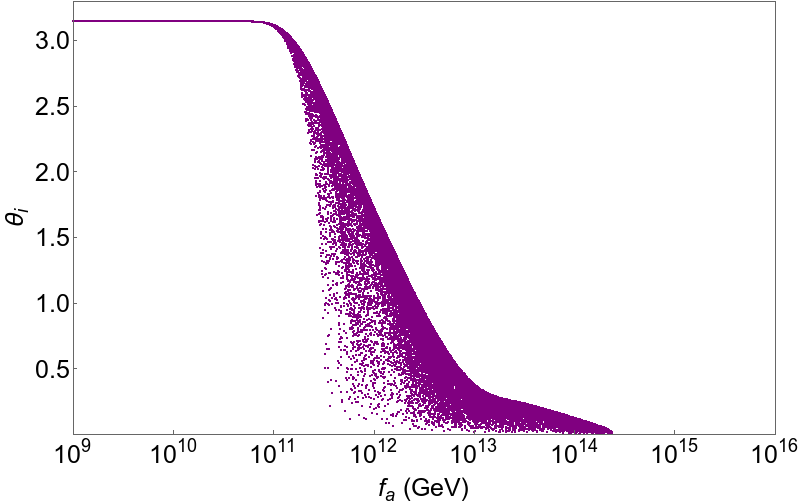}}
\caption{In {\it a}), we plot the relic density of DFSZ axions and 
higgsino-like WIMPs from a natural SUSY 
benchmark model using a scan over PQMSSM parameters in the 
SUSY DFSZ axion model.
In {\it b}), we plot the required value of axion misalignment angle 
$\theta_i$ vs. $f_a$ such that the calculated relic density 
of mixed higgsino-axion dark matter
matches the measured value $\Omega_{a\tz_1}h^2=0.12$.
\label{fig:relic}}
\end{center}
\end{figure}

The mixed axion-higgsino abundance results vs. 
$f_a$ are shown \footnote{In accord with the Particle Data Book~\cite{pdb}, we take $f_a\equiv f_A\times N_{\rm DW}$
where the domain-wall number $N_{\rm DW}=6$ for the DFSZ model.} 
in Fig. \ref{fig:relic}{\it a}).
The blue dots show the abundance of higgsino-like WIMPs while the green dots 
show the relic abundance of axions. Blue contour lines show the borders of the WIMP abundance in the allowed region. Red-colored points are excluded by
BBN constraints~\cite{jedamzik} which here occur at very large $f_a$ values where
saxions are produced at large rates via COs, but then decay late due to
couplings suppressed by large $f_a$. 
The brown points are also excluded due to $s\to aa$ decay which feeds 
relativistic degrees of freedom into the cosmic plasma so that 
$\Delta N_{\rm eff}>1$~\cite{planck}. 
For very low values of $f_a$, the
higgsino-like WIMPs are at their thermal abundance value since 
axinos and saxions decay well before neutralino freeze-out. 
The axions make up the bulk of DM in this case, at the 95\% level. 
For very low $f_a$, this seems artificial since $\theta_i$ can be adjusted to very nearly $\pi$ and the axion field would have to sit close to the maximum of its potential. This can be seen from frame {\it b}) where we show the value 
of $\theta_i$ which is required to enforce $\Omega_{a \tz_1}h^2=0.12$ 
vs. $f_a$. 

As $f_a$ increases to nearly $10^{11}$ GeV, then some axinos start decaying 
after neutralino freeze-out thus generating in addition a 
non-thermal population of WIMPs. 
If enough WIMPs are injected via decays, then the WIMPs may 
re-annihilate at the (lower) axino (or saxion) decay temperature 
which still leads to an increased non-thermal WIMP abundance~\cite{az1}.
For some parameter space points, the non-thermal WIMP production pushes the WIMP
abundance above the meaured value $\Omega_{\tz_1}h^2>0.12$ 
and so the points become excluded. Also, for $f_a\agt 10^{11}$ GeV, the required
axion misalignment angle $\theta_i$ begins dropping into a more realistic 
range. 

For yet higher values of $f_a\agt 10^{13}$ GeV, the minimal
WIMP relic density begins decreasing (somewhat buried beneath the green axion points). This is due to increasing CO-production of saxions followed by 
their late decays. In the case shown here, the saxion couples to 
axions and axinos via
\be
{\cal L}\ni\frac{\xi_s}{f_a}s\left[ (\partial_\mu a)^2+i\bar{\ta}\dsl\ta\right]
\ee
where the factor $\xi_s$ denotes the model dependence of the 
$saa$ and $s\ta\ta$ couplings~\cite{cl}. 
The value $\xi_s$ can vary between $0-1$ 
with perhaps some theory prejudice for $\xi_s\sim 1$ (which we adopt here).
For $\xi_s\sim 1$, then the saxion dominantly decays to $aa$ or if kinematically allowed, at comparable rates into $\ta\ta$. The $s$ may also decay into SUSY 
particles and SM particles-- complete decay rates and example branching fractions are shown in Ref.~\cite{dfsz1}. Now back to the dip in Fig. \ref{fig:relic}{\it a}). These dip points occur for parameter values with $m_s<2m_{\ta}$ 
where saxion decays as $s\to aa$ dominantly but also 
with substantial $s\to {\rm SM\ particles}$. 
The latter decays inject entropy into the cosmic plasma such as to dilute all relics present. 
Thus, the dip occurs because of saxion decay induced entropy dilution.
For yet higher values of $f_a$, the minimal value of $\Omega_{\tz_1}h^2$ 
turns up again as the delayed saxion decay into WIMPs wins out over entropy dilution. For $f_a\agt 10^{14}$ GeV, then always too much WIMP DM is produced
and the value of $f_a$ is disallowed. Actually, this very high $f_a$ range
becomes triply disallowed because also saxions begin decaying after the onset 
of BBN, thus violating limits on late-decaying neutral cosmic 
particles~\cite{jedamzik} (red points) 
and also too many relativistic axions are injected
into the cosmic plasma thus violating CMB bounds on extra species of 
relativistic particles (parametrized in terms of limits on the effective additional neutrino species~\cite{pdb,planck} 
which we take conservatively as $\Delta N_{\rm eff}>1$ (brown points)).

For $f_a\agt 10^{12}$ GeV, then always the WIMP abundance is 
non-thermally elevated and many more points are excluded by generating 
too much WIMP dark matter. A new set of constraints now also impact the
PQMSSM parameter space: as the WIMP abundance increases, then the
fraction of WIMP dark matter $\xi\equiv \Omega_{\tz_1}h^2/0.12$ increases
and even though $\xi<1$, the WIMPs may come into conflict with
spin-independent ({\bf SI}), spin-dependent ({\bf SD}) 
and indirect WIMP detection ({\bf IDD})
constraints~\cite{wimp}.  
It is worth noting that we assume the WIMP fraction $\xi$ is the same in the whole universe.
\begin{figure}[tbp]
\begin{center}
\hspace{-0.15cm}\includegraphics[height=0.293\textheight]{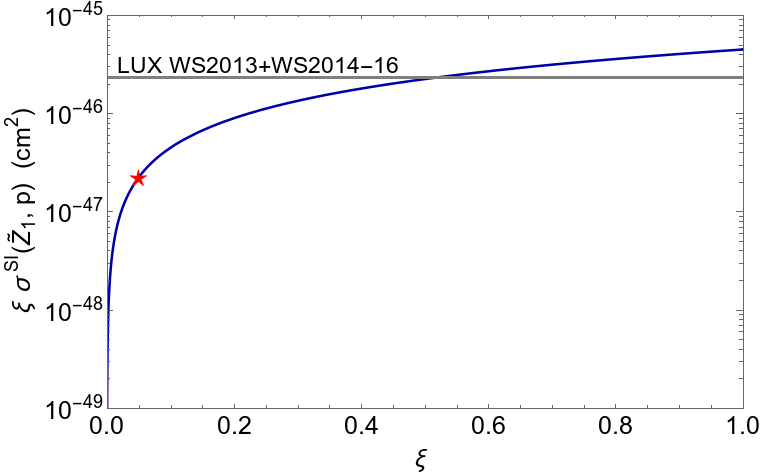}\\
\includegraphics[height=0.29\textheight]{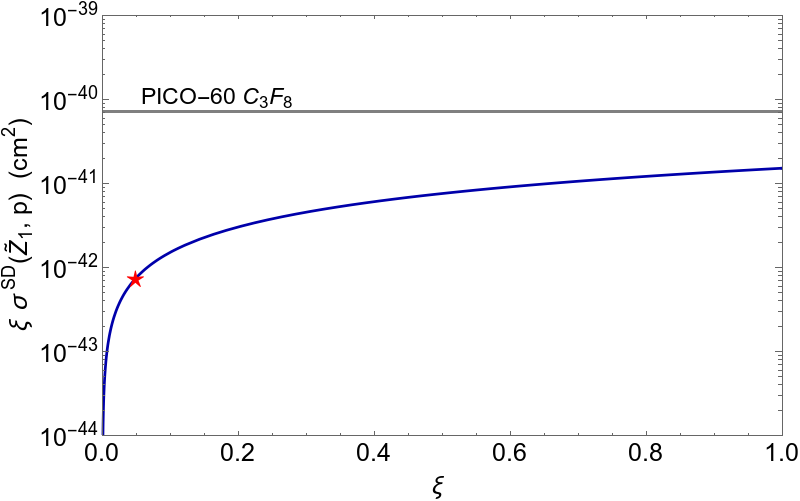}\\
\includegraphics[height=0.29\textheight]{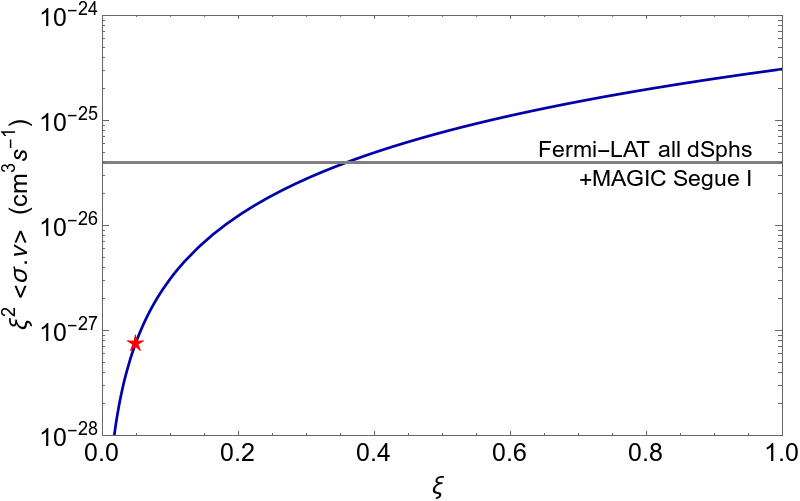}
\caption{In {\it a}), we plot $\xi\sigma^{\rm SI}(\tz_1,p)$ vs. 
$\xi\equiv \Omega_{\tz_1}h^2/0.12$ for our natural SUSY benchmark point. 
The red star denotes the value of $\xi$ obtained from thermal WIMP 
production only. The horizontal line denotes the upper limit 
reported from the LUX experiment.
In {\it b}), we plot $\xi\sigma^{\rm SD}(\tz_1,p)$ along with the
upper bound from PICO and in {\it c}) we plot
$\xi^2\langle\sigma v\rangle$ with an upper limit from Fermi-LAT/MAGIC
searches for $\gamma$ emissions from dwarf spheroidal galaxies.
\label{fig:wimp}}
\end{center}
\end{figure}
In Fig. \ref{fig:wimp}, we plot in {\it a}) the value of 
$\xi\sigma^{\rm SI}(\tz_1,p )$ and {\it b}) $\xi\sigma^{\rm SD}(\tz_1,p )$ versus
the value of $\xi$ and in {\it c}) we plot $\xi^2\langle\sigma v\rangle$ versus
$\xi$. From frame {\it a}), we see that as $\xi$ increases, the value 
of $\xi\sigma^{\rm SI}(\tz_1,p )$ approaches and then exceeds the latest constraint 
from the LUX experiment~\cite{lux}. The locus of the thermal value of
$\xi$ for our benchmark point is denoted by a red star. Once $\xi >0.53$, 
then the parameter space point violates the LUX limit on SI direct WIMP 
detection. From frame {\it b}), we see that as $\xi$ increases, then the
value of $\xi\sigma^{\rm SD}(\tz_1,p )$ always stays below the most constraining
SD limit which is currently from the PICO hot liquid bubble 
experiment~\cite{pico}. In frame {\it c}), we find that as $\xi$ increases, 
now the value of $\xi^2\langle\sigma v\rangle$ exceeds limits from
Fermi-LAT/MAGIC searches~\cite{fermi} 
for gamma-ray emissions from WIMP-WIMP annihilation to $W^+W^-$
(this annihilation channel is most relevant for higgsino-like WIMPs) for a value
of $\xi\agt 0.35$. Thus, frame {\it c}) in IDD offers the most
constraining limit on $\xi$. 

\begin{figure}[h]
  \centering
  \begin{tabular}[b]{c}
    \includegraphics[width=.471\linewidth]{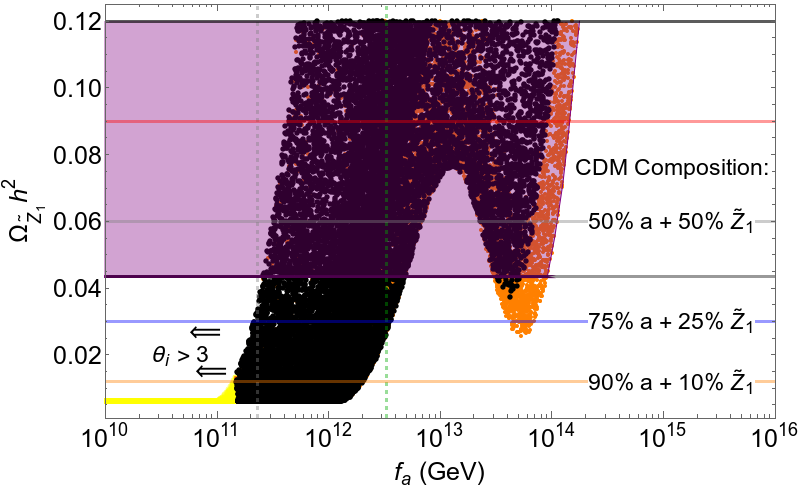} \\
    \small (a)
  \end{tabular}
  \begin{tabular}[b]{c}
    \includegraphics[width=.471\linewidth]{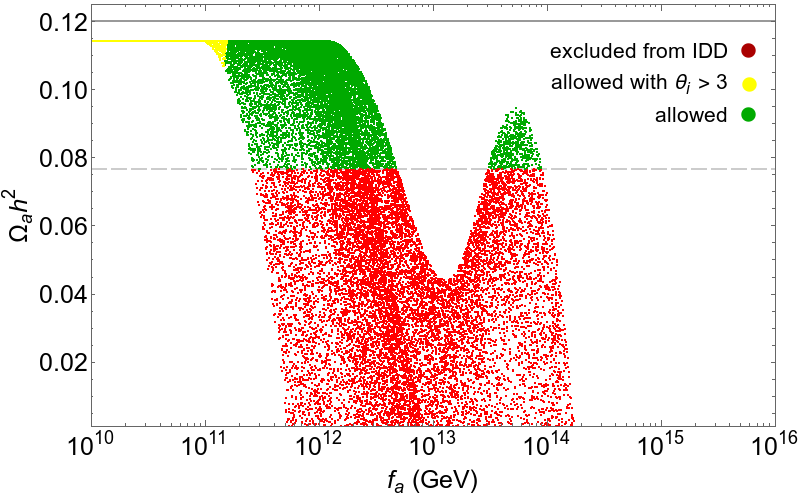} \\
    \small (b)
  \end{tabular}
  \caption{Relic density of (a) higgsino-like WIMPs and 
(b) axions from a natural SUSY benchmark model with a scan over 
PQMSSM parameters in the SUSY DFSZ axion model.
In (a), the purple shaded region is excluded by Fermi-LAT/MAGIC limits on
WIMP annihilation to gamma ray production in dwarph spheroidal galaxies.
Black dots denote where $1\text{ TeV}<m_s<30\text{ TeV}$ while orange dots denote where
$30\text{ TeV}<m_s<40\text{ TeV}$ which enhances entropy dilution from saxion decay.
%Points below purple-shade band are allowed.
Yellow dots denote where $\theta_i>3$.
}
\label{fig:relic2}
\end{figure}

An overview of the ultimate allowed region of $f_a$ in the natural SUSY DFSZ
model is shown in Fig. \ref{fig:relic2}{\it a}) where we present a blown-up version of Fig. \ref{fig:relic}{\it a}). In this case, the purple-shaded region denotes
values of $\Omega_{\tz_1}h^2$ which are in violation of the Fermi-LAT/MAGIC dwarph-spheroidal constraint. The black points (scanning over $1\text{ TeV}<m_s<30\text{ TeV}$) with $\Omega_{\tz_1}h^2\alt 0.45$ are
thus fully allowed. The yellow points are seemingly allowed as well, but these
points require $\theta_i>3$ and seem rather implausible. 
The allowed region of $f_a$ thus breaks up into two disjoint bands: 
the first runs from $10^{11}$ GeV$<f_a <4\times 10^{12}$ GeV. 
Then a gap for values $4\times 10^{12}$ GeV$<f_a <3\times 10^{13}$ GeV
ensues where WIMP production is sufficiently large that Fermi-LAT/MAGIC 
constraints dis-allow the parameter space. For
$3\times 10^{13}$ GeV$<f_a <10^{14}$ GeV, then the $f_a$ values are
re-allowed due to the effect of saxion entropy dilution of the WIMP
abundance (and where WIMP production via $s\to \ta\ta$ is kinematically forbidden). The orange colored points denote where $m_{s}\agt 30$ TeV showing that
most of this region comes from very heavy saxions. For heavy enough
saxions, then the saxion decay rate is enhanced which helps avoid
WIMP overproduction by late decays. For even higher $f_a>10^{14}$ GeV 
values, then all of parameter space is disallowed.

In Fig. \ref{fig:relic2}{\it b}), we present the axion density so that $\Omega_ah^2+\Omega_{\tz_1}h^2=0.12$. Red dots show the axion density which satisfy 
the relic density constraint along with WIMPs but excluded from the indirect searches for overestimating the WIMP abundance (correspond to the purple-shaded region 
in Fig. \ref{fig:relic2}{\it a})). Allowed 
parameter space is shown by the green dots where axions make up more than 75\% of the total dark matter density.

\section{Axion detection at microwave cavity experiments}
\label{sec:axion}

Next, we address the prospects for axion detection in the 
natural SUSY DFSZ model. At present, the most sensitive experiment searching 
for QCD axions is the Axion Dark Matter search eXperiment, 
or ADMX~\cite{admx,rybka}. ADMX implements a super-cooled microwave cavity 
(Sikivie) detector~\cite{sikivie} which can be tuned over a range of frequencies
to search for axion-to-photon conversion in the presence of a strong 
$\vec{B}$-field. 
The power produced in the cavity at a frequency $\nu_a$ corresponding to
an axion mass $m_a$ is given by
\be
P=g_{a\gamma\gamma}^2\frac{\rho_a}{m_a}B_0^2 V C_{mnp} Q_L
\label{eq:P}
\ee
where $g_{a\gamma\gamma}$ is the (model dependent) axion-photon-photon coupling,
$\rho_a$ is the axion local density, $B_0$ is the magnetic field strength,
$V$ is the volume of the cavity and $Q_L$ is the loaded quality factor of the cavity. $C_{mnp}$ denotes a normalized coupling form factor of the axion to a
specific frequency mode. 
The relevant quantities of theoretical interest are then the axion mass $m_a$, 
the axion local density $\rho_a$ and the 
axion coupling $g_{a\gamma\gamma}$. 
%Whereas usually $\rho_a$ is assumed to give
%$\Omega_ah^2\equiv (\rho_a /\rho_c)h^2=0.12$ (where $\rho_c$ is the critical 
%closure density), 
Whereas usually the axion local density is taken to be 
the same as the DM local density,
in our case since axions only make up a portion of the dark 
matter: then we define $\xi_a\equiv \Omega_ah^2/0.12$ which gives the fractional
axion dark matter density around the earth.
Also, we remind the reader that~\cite{axionmass}
\be
m_a=\frac{z^{1/2}}{1+z}\frac{f_\pi m_\pi}{(f_a/N_{\rm DW})}
\simeq 6\mu{\rm eV}\frac{10^{12}\ {\rm GeV}}{(f_a/N_{\rm DW})} 
\ee
where $z=m_u/m_d\simeq 0.56$ but with a plausible range $z:0.38-0.58$~\cite{pdb}.

The axion coupling strength is given by~\cite{axreview,pdb,kaplan1985,srednicki}
\be
g_{a\gamma\gamma}=\frac{\alpha}{2\pi (f_a/N_{\rm DW})}\left(\frac{E}{N}-
\frac{2}{3}\frac{4+z}{1+z}\right)
\label{eq:g_agmgm}
\ee
where $E$ and $N$ are the electromagnetic and color anomalies
of the axion axial current. 
\begin{figure}[tbp]
\begin{center}
\includegraphics[height=0.2\textheight]{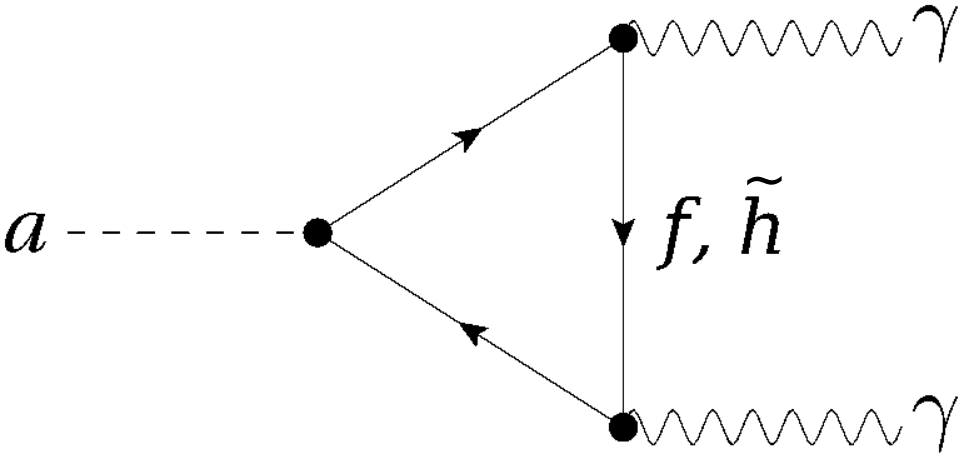}
\caption{Feynman diagram showing the fermionic loop contributions to the 
$a-\gamma -\gamma$ coupling.
\label{fig:diagram}}
\end{center}
\end{figure}

The second contribution in parenthesis of Eq. \ref{eq:g_agmgm} (from 
chiral symmetry) gives a value $\sim -1.96$ for $z=0.56$. 
The anomaly contribution $E/N$ arises from fermions circulating in 
Fig.~\ref{fig:diagram}: $E=\sum_{\rm fermions}Q_{\rm em}^2\times n_{\rm col}\times n_{\rm gen}\times Q_{\rm PQ}$. 
For the non-SUSY KSVZ axion model with uncharged
($Q_{\rm em}=0$) heavy color triplets circulating, then $E/N=0$
so that $g_{a\gamma\gamma}^{\rm KSVZ}\simeq -1.96 \alpha/2\pi (f_a/N_{\rm DW})$.
For the non-SUSY DFSZ model, then $Q_{\rm PQ}(Q,L)=+1$, $Q_{\rm PQ}(U^c,D^c,E^c)=0$
so for ($u,c,t)$ quarks we obtain $E=4$ and for $(d,s,b)$ quarks we obtain
$E=1$ and for $e,\mu,\tau$ leptons we obtain $E=3$. 
The sum yields $E/N=8/3$ or 
$g_{a\gamma\gamma}^{\rm DFSZ}\simeq 0.7\alpha/2\pi (f_a/N_{\rm DW})$. 
For our case of the SUSY DFSZ axion model, we must also add in the higgsino
contribution to $g_{a\gamma\gamma}$ with $Q_{\rm PQ}(\widetilde{H}_u,\widetilde{H}_d)=-1$ so that
$E_{\tilde{h}}=-2$. Summing over fermions yields $E/N=6/3$ so that there is a near
cancellation between the anomaly and chiral symmetry contributions:
$g_{a\gamma\gamma}^{\text{SUSY DFSZ}}\simeq 0.04\alpha/2\pi (f_a/N_{\rm DW})$ for $z=0.56$.
In fact, if $z$ is identically $0.5$, 
then an  exact cancellation occurs and the coupling drops to zero
(such a diminished $a\gamma\gamma$ coupling has been noted previously 
for non-SUSY models in Ref's~\cite{kaplan1985}). 
In such a case, apparently the invisible axion becomes again 
{\it invisible} with respect to its coupling to photons.

\begin{figure}[tbp]
\begin{center}
\includegraphics[height=0.45\textheight]{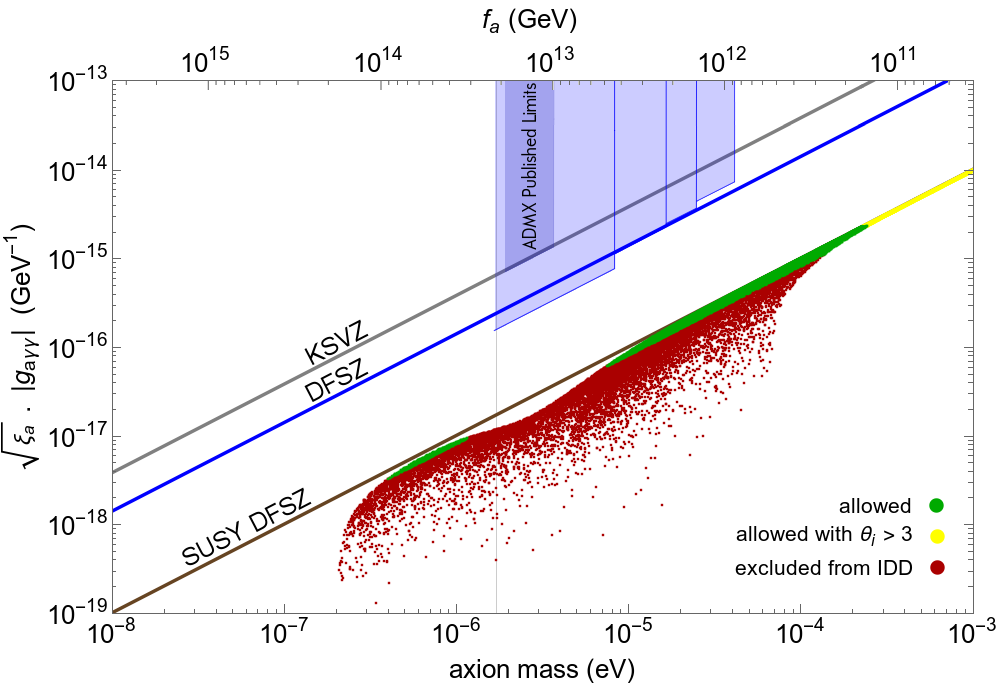}
\caption{Axion detection rates at microwave cavity experiments in terms of the
axion coupling $|g_{a\gamma\gamma}|$ vs. $m_a$. The vertical axis includes a factor
$\sqrt{\xi_a}$ where $\xi_a\equiv \Omega_ah^2/0.12$ to account for the depleted abundance of axions.
The green points are allowed from natural SUSY while red points are excluded by Fermi-LAT
constraints on higgsino-like WIMP annihilation into gamma rays. We also plot lines of 
SUSY and non-SUSY coupling strengths and current and projected ADMX search regions.
The yellow dots are regarded as unnatural since they would require an
axion misalignment angle $\theta_i>3$.
\label{fig:axion}}
\end{center}
\end{figure}

The non-SUSY KSVZ, non-SUSY DFSZ and SUSY DFSZ coupling are shown in 
Fig. \ref{fig:axion} in the $m_a$ vs. 
$\sqrt{\xi_a}\cdot |g_{a\gamma\gamma} |$ plane. We also show the ADMX published 
limits~\cite{rybka} in dark blue and ADMX proposed future search region 
as light blue. The latter extends down to the non-SUSY DFSZ coupling strength.
We also show the value of $\xi_a\cdot |g_{a\gamma\gamma} |$ obtained from 
our natural SUSY benchmark point with a scan over SUSY DFSZ parameters.
The green dots show allowed points whilst the red points are excluded
by IDD WIMP search results from Fermi-LAT. 
These latter points have a diminished axion abundance since the
WIMP abundance is non-thermally-enhanced. The green allowed points
actually break into two disjoint regions: the lower $m_a$ region
corresponds to large $f_a$ where CO-produced saxions bring the higgsino-like
WIMP abundance into accord with the measured DM relic density via 
entropy dilution. The larger $m_a$-allowed region corresponds to the 
$f_a\sim 10^{11}$ GeV-$3\times 10^{12}$ GeV region of Fig's \ref{fig:relic} and
\ref{fig:relic2}. The current ADMX search region lies in the intermediate
disallowed region where thermal axino production and decay contributes to
non-thermal higgsino-like WIMP production which lies in the IDD-constrained
region. The SUSY DFSZ coupling strength lies a factor 17 below the 
non-SUSY DFSZ projection. 
The non-SUSY DFSZ line has recently been reached in actual 
ADMX searches~\cite{tanner}.
The yellow points at very large $m_a>3\times 10^{-4}$ eV seem implausible 
due to some tuning required for $\theta_i\simeq \pi$.

A new detection method has been proposed in Ref.~\cite{Kahn:2016aff} that is sensitive to the QCD axion mass ($m_a \sim [10^{-14},10^{-6}]$ eV) 
predicted by string/GUT inspired axion dark matter models. Although the proposed experiment is projected to probe $g_{a\gamma\gamma}$ down to 
$10^{-19}$ GeV$^{-1}$, the targeted parameter space does not probe our allowed
region.

\section{Conclusions} 
\label{sec:conclude}

In this paper we have reported on prospects for axion detection in a 
model which allows for 
\bi 
\item naturalness in the EW sector via SUSY and 
\item naturalness in the QCD sector via inclusion of PQ symmetry 
and its concommitant axion.
\ei
We specialize to the SUSY DFSZ axion model since this allows for both 
\bi
\item a solution to the SUSY $\mu$ problem (why the superpotential $\mu$ 
parameter is weak scale rather than Plank scale) and 
\item allows for the required Little Hierarchy
$\mu\sim 100-300\ {\rm GeV}\ll m_{\rm SUSY}\sim 1-10$ TeV.
\ei
A subset of these models actually allows for radiative PQ breaking 
where the PQ scale $f_a$ is generated from SUSY breaking. Such models
tend to generate a value $\mu\sim 100-300$ GeV from soft SUSY breaking 
terms of order $1-10$ TeV~\cite{radpq}. For these reasons, we feel the model
explored in this paper is the most highly motivated axion model available 
so that the results presented here should be regarded more as a 
paradigm of what is to be expected for axion physics, 
rather than some implausible outlier model.

Our main result is summarized in Fig. \ref{fig:axion}. This plot shows a 
wide range of $m_a$ values over which an axion might be expected. 
The two allowed regions are disjoint: 
the upper region with $m_a:10^{-5}-3\times 10^{-4}$ eV occurs with mainly axion
cold dark matter and a smaller contribution of higgsino-like WIMPs. 
It is somewhat higher in $m_a$ values than is currently being explored at ADMX.
The lower region with $m_a:3\times 10^{-7}-1.5\times 10^{-6}$ eV 
occurs due to CO production of (very heavy) saxions followed by decays to 
SM (and other) particles which leads to entropy dilution of all relics. 
An intermediate region is actually excluded by Fermi-LAT bounds on
WIMP-WIMP annihilation into gamma rays where the WIMP abundance is non-thermally enhanced mainly due to axino/saxion production and decay in the early universe.

What is more distressing is that in the PQ augmented MSSM, then the axion coupling is severely depressed relative to non-SUSY KSVZ or DFSZ via the inclusion 
of higgsinos in the $g_{a\gamma\gamma}$ triangle coupling. For $z=m_u/m_d=0.56$, then the coupling is suppressed by a factor around 17 although the coupling could be even more suppressed for $z=0.5$. 
The detection rate is somewhat reduced as well due to the fact that 
axions only make up a portion of the dark matter abundance. This necessitates
inclusion of a factor $\sqrt{\xi_a}$ in the axion coupling vs. $m_a$ plot.

While our results were presented for just one SUSY benchmark, 
it should be noted that they are still rather general since naturalness 
requires $\mu\sim 100-300$ GeV (the closer to $m_Z$ the better) while 
LHC sparticle mass limits and Higgs mass measurement require 
sparticle masses, especially gluinos and squarks, in the multi-TeV range. 
Generally, all natural models based on the MSSM should look pretty close
to our benchmark as far as dark matter physics is concerned. 

Ultimately, these results seem to show the axion may exist across a broader
mass range than otherwise might be expected. 
Even so in a SUSY DFSZ scenario, axions from stringy models~\cite{ws,bl} 
with $f_a\sim 10^{16-18}$ GeV 
seem highly unlikely in this regard\footnote{An exception occurs when saxion decays to neutralinos are not kinematically allowed and the $\mu$ term 
is large so saxions decaying into gauge bosons and into the Higgses are the dominant decay modes~\cite{dfsz1,dfsz2}.} since saxion 
production via COs and very late decay makes model points triply excluded via:
1. over non-thermal-production of WIMPs, 2. injection of relativistic particles
($s\to aa$) into the cosmic plasma (violating bounds on $\Delta N_{\rm eff}$)
and 3. violations of BBN constraints from late-time saxion decays.
Also, our results show the axion signal strength may be far lower than 
might be expected from non-SUSY axion models due to inclusion of higgsinos
circulating in the $a\gamma\gamma$ triangle loop. Thus, axions may be rendered once again more invisible to experiment than otherwise anticipated. 
As a result, we urge new ideas and initiatives to probe more broadly and more
deeply into the $g_{a\gamma\gamma}$ vs. $m_a$ axion parameter space.

\section*{Acknowledgments}

We thank Jihn E. Kim and Seokhoon Yun for discussion and Vernon Barger 
for a close reading of this manuscript. 
This work was supported in part by the US Department of Energy, 
Office of High Energy Physics. 
The work of KJB was supported by IBS under the project code, IBS-R018-D1.
The computing for this project was performed at the 
OU Supercomputing Center for Education \& Research (OSCER) 
at the University of Oklahoma (OU).

%
%%%%%%%%%%%%%%%%%%%%%%%%%%%%%%%%%%%%%%%%%%%%%%%%%%%%%%

%
\end{document}